\documentclass[conference]{IEEEtran}

\usepackage{epsfig}
\usepackage{times}
\usepackage{float}
\usepackage{afterpage}
\usepackage{amsmath}
\usepackage{amstext,cite}
\usepackage{amssymb,bm}
\usepackage{latexsym}
\usepackage{color}
\usepackage{graphicx}
\usepackage{amsmath}
\usepackage{amsthm}
\usepackage{graphicx}
\usepackage[center]{caption}
\usepackage{pstricks}
\usepackage{caption}
\usepackage{subcaption}
\usepackage{booktabs}
\usepackage{multicol}
\usepackage{lipsum}% dummy text
\usepackage[T1]{fontenc}
\usepackage{hyperref}
 \usepackage{aecompl}
 \usepackage{mathrsfs}

\usepackage{soul}
\usepackage{cancel}

\usepackage{changes}
\definechangesauthor[color=red,name={Mingyue Ji}]{MJ}

\allowdisplaybreaks

\long\def\comment#1{}

\newcommand{\Mc}{{\mathcal M}}
\newcommand{\Nc}{{\mathcal N}}

\newcommand{\Sc}{{\mathcal S}}

\newcommand{\Wc}{{\mathcal W}}

\newcommand{\asf}{{\mathsf a}}

\newcommand{\csf}{{\mathsf c}}

\newcommand{\esf}{{\mathsf e}}

\newcommand{\ssf}{{\mathsf s}}

\newcommand{\wsf}{{\mathsf w}}

\newcommand{\Ksf}{{\mathsf K}}

\newcommand{\Nsf}{{\mathsf N}}

\newcommand{\Qsf}{{\mathsf Q}}

\newcommand{\Tsf}{{\mathsf T}}

\renewcommand{\arg}{{\hbox{arg}}}

\newtheorem{thm}{Theorem}%[section]

\providecommand{\definitionname}{Definition}

\begin{document}

\title{Topological Coded Distributed Computing} 
\author{
\IEEEauthorblockN{%
Kai~Wan\IEEEauthorrefmark{1}, 
Mingyue~Ji\IEEEauthorrefmark{2}, 
Giuseppe~Caire\IEEEauthorrefmark{1}
}
\IEEEauthorblockA{\IEEEauthorrefmark{1}Technische Universit\"at Berlin, 10587 Berlin, Germany, \{kai.wan, caire\}@tu-berlin.de}% 
\IEEEauthorblockA{\IEEEauthorrefmark{2}University of Utah, Salt Lake City, UT 84112, USA, mingyue.ji@utah.edu}% 
}

\maketitle
%\IEEEpeerreviewmaketitle{}

\begin{abstract}
This paper considers the MapReduce-like coded distributed computing framework originally proposed by Li et al., 
which   uses coding  techniques when %some  
distributed computing servers  exchange their computed intermediate values, in order to reduce the overall traffic load. %which is considered as one of the main bottleneck in the MapReduce-like applications. %number of transmissions. 
Their original model servers are connected via an error-free common communication bus allowing broadcast transmissions. %such that the traffic load can be directly translated to the communication delay.  
However, this  assumption is one of the major limitations in practice since the practical cloud computing network topologies are far more involved than a simple single bus.    % and the overall traffic load may not be directly translated to the communication delay. 
 We formulate a topological coded distributed computing problem, where the distributed servers communicate with each other through some switch network. 
By using a special instance of fat-tree topologies, referred to as {\it $t$-ary fat-tree} proposed by Al-Fares et al.  which can be built by  some cheap switches, we propose a coded distributed computing scheme to achieve the minimum max-link communication load defined as the maximum load over all links. %communication delay.  
\end{abstract}


\section{Introduction}
\label{sec:intro}
 
% {\magenta Comments of Kai: some background on distributed computing is missing.}

Recent years have witnessed the emergence of big data with wide range of applications. %in both business and consumer worlds. 
 To cope with such a large dimension of data and the complexity of data mining algorithm, it is increasingly popular to use cloud computing platforms such as Amazon Web Services (AWS) \cite{amazon2015amazon}, Google Cloud \cite{gcp2015}, and 
Microsoft Azure \cite{wilder2012cloud}. In particular, modern distributed computing platforms such as MapReduce \cite{dean2008mapreduce} and Spark \cite{zaharia2010spark} have attracted significant attentions since they enable the computation of large tasks on data sizes of order of terabytes. The path to exascale distributed computing poses a number of significant challenges as researchers and potential exascale vendors attempt to deliver a hundred times performance improvement relative to today's distributed computing systems. While large scale distributed algorithms and simulations running at these extreme scales have the potential for achieving unprecedented levels of accuracy and providing dramatic insights into complex phenomena, they are also presenting new challenges. Keys among these are the challenges related to the computation and communication costs. 
In order to tackle these large-scale problems,  it is critically important to understand the fundamental tradeoff between computation and communication. Inspired from the idea from the current development of coded caching networks \cite{dvbt2fundamental,d2dcaching}, the pioneer works \cite{li2015coded,distributedcomputing} introduces the concept of {\em Coded Distributed Computing} (CDC), which enables network coding among intermediate computed values to save significant communication load among servers. %Based on this idea, 
In particular, \cite{distributedcomputing} studied the fundamental tradeoff between communication load and computation load in a ``MapReduce-like" distributed computing system. Surprisingly, in %the information theoretical sense, 
theory, it showed that if a task can be computed repeatly at $r$ workers, the total communication load $L(r)$ can potentially be reduced $r$ times. This means that we can trade computation power for communication load, which has the potential to lead to a solution of the %communication load 
traffic congestion problem in the current distributed computing systems.

%The seminal paper by Li et al.~\cite{distributedcomputing} proposed a MapReduce-based coded distributed computing framework, which 
The framework considered in \cite{distributedcomputing} contains  {\it Map}, {\it Shuffle}, and {\it Reduce} phases. 
In the map phase, the distributed computing servers process parts of the stored data locally and generate some intermediate values.
In the shuffle phase, each server broadcasts some computed  intermediate values  to other servers through an error-free common-bus (each server can receive the packets transmitted by other servers without error),    such that all the servers can obtain enough input values to compute the output functions in the reduce phase. Other aspects and extensions of CDC are considered in the literature such as reducing complexity \cite{8437323,konstantinidis2019resolvable,8437882}, randomized connectivity \cite{8437653}, alternative metrics \cite{8278011}, and in wireless channels~\cite{wirelessCDC2017}.
%The coded distributed computing scheme was proved in~\cite{distributedcomputing} to  achieve the optimal tradeoff between the computation load in the map phase and the communication load in the shuffle phase. 
%  {\magenta Comments of Kai: which following works of the distributed computing should be discussed later?} 

 As pointed out in~\cite{distributedcomputing,fogcomputing}, while the common-bus topology is meaningful for co-located processors, it is generally difficult to implement such topology for physically separated servers.  Since the publications of~\cite{distributedcomputing,fogcomputing}, designing a practical data center network topology that can reap the gains of coded distributed computing in terms of per-link communication load is widely open. 
In this paper, we consider a general switch network connecting the computation servers as illustrated in Fig.~\ref{fig: system_model}.  Depending on the network
topology, the  max-link communication load may be
more relevant compared to the total communication load sent
from each server.  Our objective is to find a practically used network topology and design appropriate coded distributed computing schemes, such that given the computation load in the map phase,
 the max-link communication load  over all links in this topology (related to the communication delay) is minimized.  
In addition to the problem formulation, our main contributions in this paper are 
\begin{itemize}
\item We characterize the optimal max-link communication load by proposing a cut-set converse  and   a coded distributed computing scheme on a single-switch topology.
\item On the observation that the cost to build the single-switch topology is high (because we need a giant switch whose number of ports should be equal to the number of servers), we propose to use   the $t$-ary fat-tree topology proposed in~\cite{scalablenetwork} as illustrated in Fig.~\ref{fig: fat tree} (detailed description on the topology will be provided in Section~\ref{sub:fat tree}), which is existing and  has been widely used in practice for the data center networks~\cite{surveyonfattree2017}.
This $t$-ary fat-tree is built by some $t$-ports switches which can handle up to $\frac{t^3}{4}$ servers, which can significantly reduce the network building cost. By leveraging {\bf the symmetry of the network and the fact that there exists some path between any two servers},  we  then propose a coded distributed computing scheme based on this  $t$-ary fat-tree, which can actually  achieve the optimal max-link communication load.
\end{itemize} 
 In one word, we are the first to show that the whole coded distributed computing gain in~\cite{distributedcomputing} can be achieved in another and practical topology rather than the oversimplified shared-link topology.  Therefore, this is an
important step in the direction of bringing coded distributed computing much closer to practice.

\paragraph*{Notation convention} 
%We use the following notation convention.
Calligraphic symbols denote sets and sans-serif symbols denote system parameters.
We use $|\cdot|$ to represent the cardinality of a set or the length of a vector;
$[a:b]:=\left\{ a,a+1,\ldots,b\right\}$ and $[n] := [1,2,\ldots,n]$.

\section{System Model}
\label{sec:model}
\subsection{Coded Distributed Computing Problem in~\cite{distributedcomputing}}
\label{sub:CDC setting}
We first briefly review the  coded distributed computing problem in~\cite{distributedcomputing}, which aims to compute $\Qsf$ arbitrary output
values (denoted by $u_1,\ldots,u_{\Qsf}$) from $\Nsf$ input files (denoted by $w_1,\ldots,w_{\Nsf}$) using a cluster of $\Ksf$ distributed  servers.  
For  some $\ssf \in [\Ksf]$ where $\binom{\Ksf}{\ssf}$ divides $\Qsf$, it is required that 
each subset of $\ssf$ servers compute a disjoint subset of  $\frac{\Qsf}{\binom{\Ksf}{\ssf}}$ output values. The set of output values which server $k$ needs to compute is denoted by $\Wc_k$. 
The computation proceeds in three phases: {\it Map}, {\it Shuffle}, and {\it Reduce}.

{\it Map phase.}
Each server $k \in [\Ksf]$  computes the Map functions of  files in $\Mc_k\subseteq [\Nsf]$, where $\Mc_k$ is stored in its memory. For each file $w_n$ where $n\in \Mc_k$,
server $k$ computes $g(w_n)=(v_{1,n},\ldots,v_{\Qsf,n})$, where $v_{q,n}$ is an intermediate value with   $\Tsf$ bits for each $q \in [\Qsf]$.\footnote{\label{foot:reduce function}The output value  $u_q$ where $q\in [\Qsf]$ can be directly computed from some intermediate values, i.e., $u_q:= h_q(v_{q,1}, \ldots , v_{q,\Nsf})$ for some function $h_q$.}
The  computation load, denoted by
 $$
r=\frac{\sum_{k\in [\Ksf]}|\Mc_k| }{\Nsf} \in [1,\Ksf],
$$
 represents the average number of nodes that map
each file.

{\it Shuffle phase.}
To compute the output value $u_q$ where $q\in \Wc_k$,  in the shuffle phase server $k$ needs to recover the intermediate values $\{v_{q,n}: n\notin \Mc_k \}$, which are not computed by itself in the map phase.
 For this purpose, each server $k$ creates an $\ell_k$-bits  message $X_k$ based on its computed intermediate values in the map phase, i.e., 
$$
X_k =\psi(\{g(w_n):n\in \Mc_k \} ).
$$
 The message $X_k$ is then broadcasted from server $k$ to other servers through a common communication bus.
The communication load, denoted by 
$$
L=\frac{\sum_{k \in [\Ksf]} \ell_k}{\Qsf \Nsf \Tsf},
$$
 represents the normalized   number of bits
communicated in the system.

{\it Reduce phase.}
 Each server $k \in [\Ksf]$ first decodes the intermediate values $\{v_{q,n}: n\notin \Mc_k \}$ from the received messages $\{X_j:j\in [\Ksf] \setminus \{k\}\}$, and then computes the output value  $u_q:= h_q(v_{q,1}, \ldots , v_{q,\Nsf})$ for each $q \in \Wc_k$.
 
 The objective is  to design the Map, Shuffle and Reduce phases such that the communication load $L^{\star}(r)$ is minimized given  the computation load $r \in [1,\Ksf]$.

 It was proved in~\cite[Theorem 2]{distributedcomputing} that the optimal tradeoff is the lower convex envelop of the following points,
 \begin{align}
 L^{\star}(r)=\sum^{\min\{r+\ssf,\Ksf\}}_{t=\max\{ r+1, \ssf\}} \frac{t\binom{\Ksf}{t}\binom{t-2}{r-1}\binom{r}{t-\ssf} }{r \binom{\Ksf}{r} \binom{\Ksf}{\ssf}},\label{eq:CDC L star}
 \end{align}
 where 
 $r\in [\Ksf]$.
 
The achievable scheme   in~\cite[Section V]{distributedcomputing} is based on linear coding. Define $T_k$ as   the transmitted message by server $k$ in the achievable scheme~\cite{distributedcomputing} for each $k\in [\Ksf]$.  $T_k$ contains $ L^{\star}(r) \Qsf \Nsf \Tsf /\Ksf$ bits, and could    be written as
a set of non-overlapping sub-messages, 
$$
T_k =\left\{T^\Sc_k: \Sc \subseteq [\Ksf]\setminus \{k\}   \right\},
$$
where $T^\Sc_k$ represents the sub-messages transmitted by server $k$ which are useful to servers in $\Sc$. 
 For each server $j \in [\Ksf]$, the set of  received sub-messages which are useful to server $j$ is defined as 
 $$
 U_j=\left\{T^\Sc_k: k\in [\Ksf]\setminus \{j\}, \Sc \subseteq [\Ksf]\setminus \{k\},j\in \Sc  \right\}.
 $$
   An important observation from the achievable scheme   in~\cite{distributedcomputing} is that  
   $$
   |U_j|=\left(\Nsf-\frac{r\Nsf}{\Ksf} \right) \frac{\Qsf\ssf\Tsf}{\Ksf},
   $$
    which is equal to the number of bits in $\{v_{q,n} : q\in\Wc_j , n \notin\Mc_j \}$ (the total length of the intermediate values to compute $u_q$ for all $q\in\Wc_j$, which are not computed by server $j$ in the map phase).
   \subsection{Topological Coded Distributed Computing}
\label{sub:topological CDC setting}
\begin{figure}%[ht]
\centerline{\includegraphics[scale=0.25]{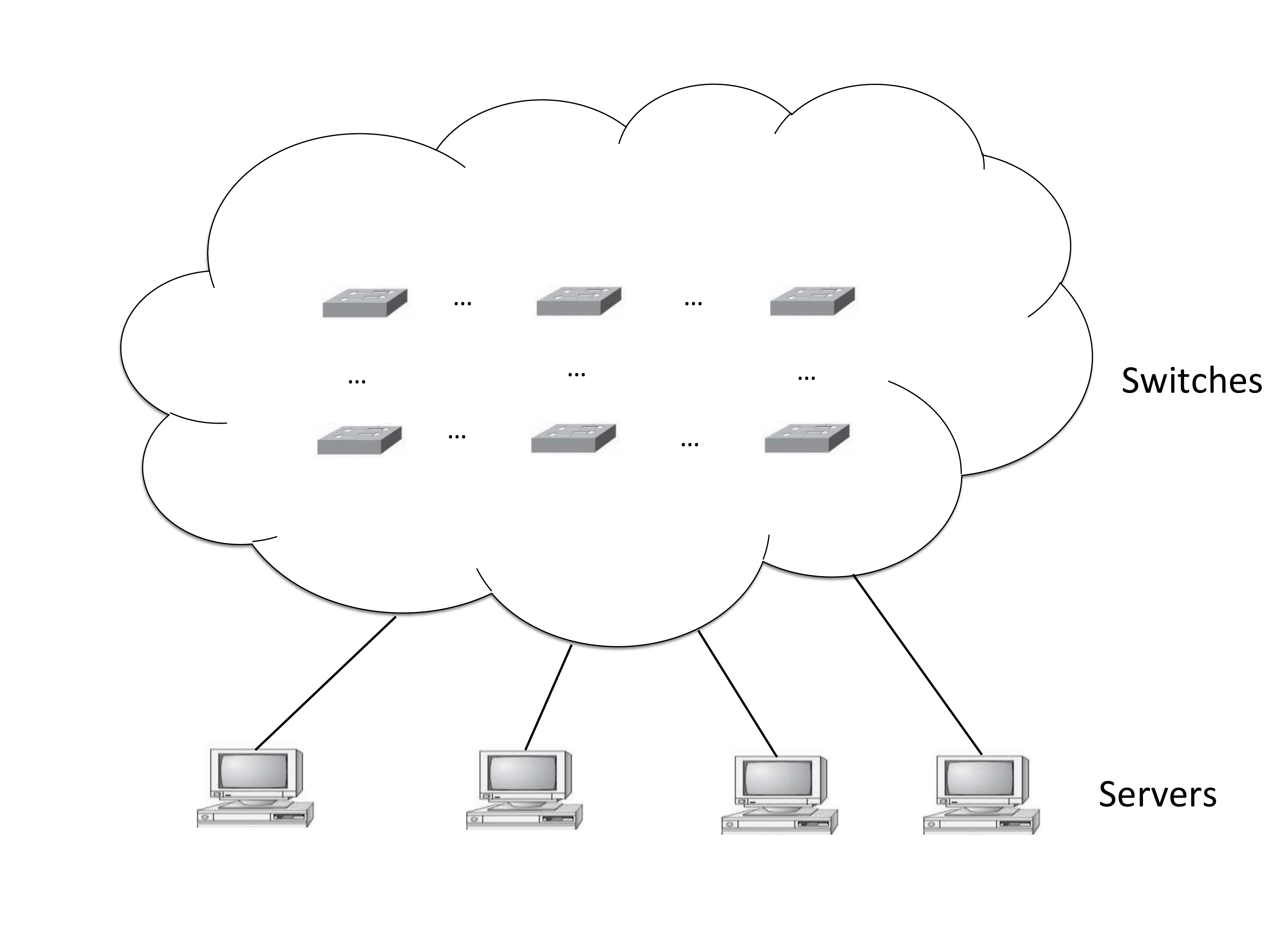}}
\caption{\small The    topological coded distributed computing problem.}
\label{fig: system_model}
\end{figure}

The coded distributed computing framework in~\cite{distributedcomputing} assumes that each server  broadcasts some message  to others through a common communication bus. However, this topology is not used in practice. %exists rarely in the practice. 
Instead, we always need to build a topological networks composed of  switches and  wired links   to enable the  communications among the servers.
Fig.~\ref{fig: system_model} illustrates the general data center networks 
%topological coded distributed computing network 
considered in this paper, where each server is connected to a cloud of switches through an individual wired link. 
The switches in the network cannot compute functions, and we assume that there does not exist any direct link from one server to another.

The map and reduce phases in our considered problem are the same as in~\cite{distributedcomputing}. In the shuffle phase, instead of assuming the common communication bus,  we  need to design the topology in the system.
  Assume there are totally $V$ wired links in the designed topology. 
For each link $v\in [V]$, the  number of uplink (i.e., from the bottom   to the top in the topology) transmitted bits through link $v$ is denoted by $R^{\text{up}}_v$, and    the  number of downlink (i.e., from the top   to the bottom) transmitted bits through link $v$ is denoted by $R^{\text{down}}_v$.  
The total number of bits transmitted through link $v$ is denoted by 
$$
R_v= R^{\text{up}}_v+R^{\text{down}}_v.
$$
We  define the max-link communication load $D$ as the maximal normalized number of bits transmitted through each link, where 
$$
D=\max_{v \in [V]} \frac{R_v}{ \Qsf \Nsf \Tsf}.
$$
The objective is to design a network topology and an achievable scheme to characterize the communication load $D^{\star}(r)$ given  the computation load $r \in [1,\Ksf]$.     %\footnote{Note that given the link throughputs, the communication delay is directly related to the max-link communication load.}

   \section{Main Results}
\label{sec:main results}
\subsection{Optimal max-link communication load}
\label{sub:optimal time}
%We provide the optimal trdeoff between the communication delay  and the the computation load
%{\RED [MJ: for the achievability, you assumed the shared link topology. So we may say it directly in the theorem.]}
\begin{thm}
\label{thm:optimal time}
For the considered topological coded  distributed computing problem, the optimal tradeoff between the max-link communication load  and the computation load  is the lower convex envelop of the following points,
   \begin{align}
 D^{\star}(r)= \frac{L^{\star}(r) }{\Ksf}+\frac{\ssf}{\Ksf}\left(1-\frac{r}{\Ksf} \right), \ \forall r \in [\Ksf]. \label{eq:CDC D star}
 \end{align}
\end{thm}
 \begin{IEEEproof}
 {\it Achievability.} 
 First, we need to design a network topology. %We consider the network where 
 We simply let all servers be connected to one switch at the top.  Recall that $T_k$ where $k\in [\Ksf]$  is the transmitted message by server $k$ in the achievable scheme~\cite{distributedcomputing}. 
  Each server $k$ transmits $T_k$ to the top switch. The switch forwards $U_j$ to each server $j \in [\Ksf]$.
  Hence, we prove that the communication load in~\eqref{eq:CDC D star} is achieved.
  
  {\it Converse.}
Denote the index of the link directly connected to server $k$ by $v_k$.
  We first consider the uplink transmission from the servers. The total uplink load through the links in $\{v_1,\ldots,v_{\Ksf}\}$ should be no less than $ L^{\star}(r)$, i.e., 
  \begin{align}
\sum_{k\in [\Ksf]} R^{\text{up}}_{v_k} \geq  L^{\star}(r). \label{eq:total uplink load}
  \end{align}
 We then consider the downlink transmission from the cloud to the servers. Recall that  in the shuffle phase server $k$ needs to recover $\{v_{q,n}: q\in\Wc_k, n\notin \Mc_k \}$, and that $|\Wc_k|=\frac{\Qsf}{\binom{\Ksf}{\ssf}} \binom{\Ksf-1}{\ssf-1}=\frac{\Qsf\ssf}{\Ksf}$. In other words, server $k$ needs to recover $\frac{\Qsf\ssf}{\Ksf} \left(\Nsf-|\Mc_k| \right)$. Hence, the total number of intermediate values needed to be recover by all servers is 
 \begin{align}
& \sum_{k\in [\Ksf]} \frac{\Qsf\ssf}{\Ksf} \left(\Nsf-|\Mc_k| \right)  %= \frac{\Qsf\ssf}{\Ksf}  \Big(\Nsf \Ksf- \sum_{k\in [\Ksf]} |\Mc_k|  \Big) \nonumber\\&
=\frac{\Qsf\ssf}{\Ksf}  \left(\Nsf \Ksf-  \Nsf r  \right). \label{eq:total downlink number of v}
 \end{align}
  Recall that the length of each intermediate value is $\Tsf$. From~\eqref{eq:total downlink number of v}, the total downlink load through the links in $\{v_1,\ldots,v_{\Ksf}\}$ can be bounded as follows,
  \begin{align}
  \sum_{k\in [\Ksf]} R^{\text{down}}_{v_k} &\geq \frac{\Qsf\ssf}{\Ksf}  \left(\Nsf \Ksf-  \Nsf r  \right) \frac{\Tsf}{\Qsf\Nsf\Tsf} \nonumber\\
  &=\ssf \left(1-\frac{r}{\Ksf}\right).\label{eq:total downlink load}
  \end{align} 
  From~\eqref{eq:total uplink load} and~\eqref{eq:total downlink load}, we have 
  \begin{align*}
 \max_{k\in [\Ksf]} R_{v_k} & \geq \frac{1}{\Ksf} \sum_{k\in [\Ksf]}( R^{\text{up}}_{v_k} + R^{\text{down}}_{v_k} )  \nonumber\\
 &\geq \frac{ L^{\star}(r)}{\Ksf}+   \frac{\ssf}{\Ksf} \left(1-\frac{r}{\Ksf}\right),
  \end{align*}
  which coincides~\eqref{eq:CDC D star}. Hence, we prove Theorem~\ref{thm:optimal time}.
 \end{IEEEproof}
 
The optimal max-link communication load could be achieved by using one switch connected to each server. However, we need a giant switch of $\Ksf$ ports, which is much more expensive than a network with small switches (see~\cite{scalablenetwork}).
One important question to ask is whether there exists a topology, with which we can achieve the optimal communication load in~\eqref{eq:CDC D star} while   the network can be constructed by low-cost switches with much fewer ports.

\subsection{Description of $t$-ary Fat-tree in~\cite{scalablenetwork}}
\label{sub:fat tree} 
\begin{figure}%[ht]
\centerline{\includegraphics[scale=0.25]{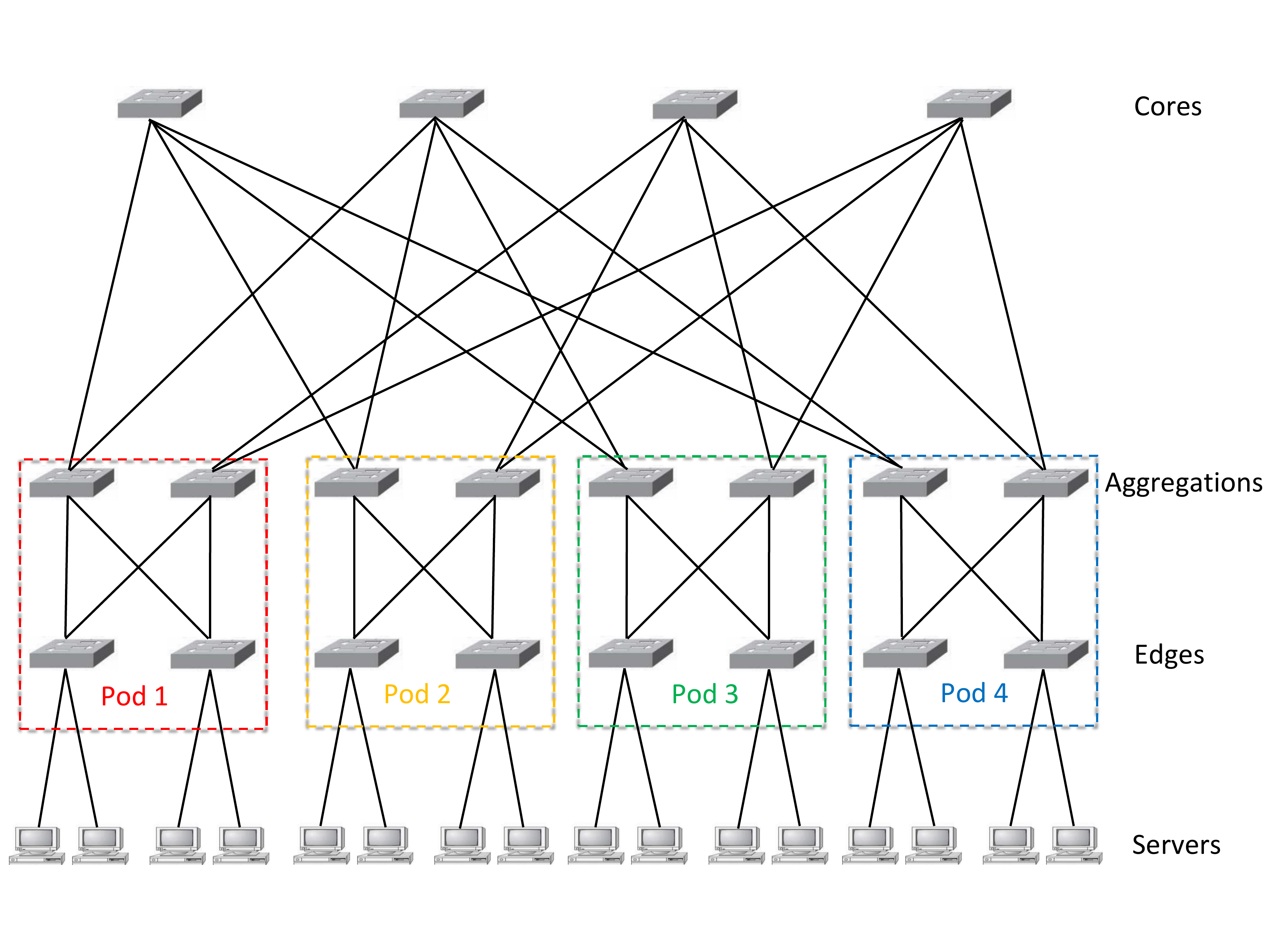}}
\caption{\small The $4$-ary fat tree.}
\label{fig: fat tree}
\end{figure}

We  can answer the question above by using  the $t$-ary fat-tree topology proposed in~\cite{scalablenetwork} (illustrated in Fig.~\ref{fig: fat tree}).
 There are four layers in the topology, with  $\frac{5 t^2}{4}$ switches in total laying in the   top three layers  and $\frac{t^3}{4}$ servers laying in the bottom layer. The switches in the top three layers are referred to as {\it cores},  {\it aggregations}, and {\it edges}, respectively, where the numbers of cores, aggregations, and edges are $\frac{t^2}{4}$, $\frac{t^2}{2}$, and $\frac{t^2}{2}$, respectively.

 The $\frac{t^2}{4}$ cores are denoted by  $\csf_{1},\ldots,\csf_{\frac{t^2}{4}}$ from  left  to  right in the network.
 A $t$-ary fat-tree topology contains $t$ pods. We focus on pod $i$ where $i\in [t]$.  Pod $i$ contains $t/2$ aggregations (denoted by $\asf_{i,1},\ldots,\asf_{i,t/2}$ from the LHS  to the RHS) and $t/2$ edges  (denoted by $\esf_{i,1},\ldots,\esf_{i,t/2}$ from left  to right). Each aggregation  $\asf_{i,j}$  where $j\in [t/2]$ is connected to $t/2$ different cores (cores $\csf_{\frac{(j-1)t}{2}+1},\ldots, \csf_{\frac{j t}{2}}$), such that each core is connected to exactly one aggregation in this pod. Aggregation $\asf_{i,j}$ is also connected to each edge in this pod. 
 Furthermore, each edge $\esf_{i,p}$ where $p\in [t/2]$ is connected to $t/2$ servers  at the bottom, and the positions of these servers are denoted by $\wsf_{i,p,1},\ldots,\wsf_{i,p,t/2}$.
 
 Hence, each switch in the fat-tree has $t$ ports, such that  $\frac{5 t^2}{4}$ $t$-ports switches can handle up to  $\frac{t^3}{4}$ servers. The cost to build this network is much cheaper than one $\frac{t^3}{4}$-ports switch.\footnote{\label{foot:cable cost}The cost of cables/links is much lower than the cost of switches. Hence, as in~\cite{scalablenetwork}, in this paper we do not consider the cost of cables.}
  %The advantage of this topology is that all switches  are identical such that cheap commodity products can be used for all switches, and that 
 
 \subsection{Coded Distributed Computing through  $t$-ary Fat-tree}
\label{sub:computing with fat tree} 
Next, we will show that with this low-cost topology, there exists an achievable scheme which can also achieve  the optimal communication load in~\eqref{eq:CDC D star}.
 The proposed achievable scheme is also based on the coded distributed computing scheme in~\cite{distributedcomputing}. The map and reduce phases are the same as the scheme  in~\cite{distributedcomputing}. In the following, we will describe how to deliver the messages $\{T_k: k\in [\Ksf]\}$ through the $t$-ary fat-tree. 
  The main intuition why the  $t$-ary Fat-tree can lead to the optimal communication load is that {\bf each edge or aggregation is connected to $t/2$ switches/servers at its lower layer and connected to $t/2$ switches at its higher layer, such that the load on each outgoing link of one switch is no more than each of its ingoing links.} 

 Based on the number of servers $\Ksf$, we choose 
 $$
 t=\arg\min_{t_1 \in \mathbb{Z}}  \frac{t_1^3}{4} \geq \Ksf.
 $$
 If $\Ksf< \frac{t^3}{4}$, 
 we  place the  $\Ksf$ servers in the first $\Ksf$ positions from the left at the bottom.  For each $i\in [t]$, $p\in [t/2]$, and $s\in [t/2]$, if there is one server (assumed to be server $k$) placed in position $\wsf_{i,p,s}$, with a slight abuse of notation, we let $\wsf_{i,p,s}=k$; otherwise, $\wsf_{i,p,s}=0$. 
In addition, we define that $\Mc_0=\Wc_{0}=T_{0}=\emptyset$.

 {\it Uplink transmission for pod $i \in [t]$.}
 \begin{itemize}
 \item Each server $\wsf_{i,p,s}$ where $p\in [t/2]$ and $s\in [t/2]$, sends $T_{\wsf_{i,p,s}}$ to its connected edge $\esf_{i,p}$. 
 
The  number of bits transmitted
through the  link from server $\wsf_{i,p,s}$ to edge $ \esf_{i,p}$ is    
\begin{align}
|T_{\wsf_{i,p,s}}| \leq L^{\star}(r)  \frac{\Qsf \Nsf \Tsf }{\Ksf}.\label{eq:up step 1 load}
\end{align} 
 \item We focus on edge   $\esf_{i,p}$ where $p\in [t/2]$. For each $s\in [t/2]$, edge  $\esf_{i,p}$  divides $T_{\wsf_{i,p,s}}$ into $t/2$ non-overlapping and equal-length pieces, denoted by $T_{\wsf_{i,p,s}}(1),\ldots,T_{\wsf_{i,p,s}}(t/2)$.
Recall that  $T^{\Sc}_{k}$ represents the sub-message in $T_k$ which are uniquely useful to servers in $\Sc$. For each $j\in [t/2]$,
we  define  $T^{\Sc}_{\wsf_{i,p,s}}(j)$ as the set of bits in $T_{\wsf_{i,p,s}}(j)$ which are uniquely useful to servers in $\Sc$. 
The above partition  of $T_{\wsf_{i,p,s}}$ is symmetric, i.e.,  
$$
\left|T^{\Sc}_{\wsf_{i,p,s}}(1)\right|=\cdots=\left|T^{\Sc}_{\wsf_{i,p,s}}(t/2)\right|=\frac{2|T^{\Sc}_{\wsf_{i,p,s}}|}{t} ,
$$ 
for each $\Sc \subseteq [\Ksf]\setminus \{\wsf_{i,p,s}\}$.
Edge   $\esf_{i,p}$ then sends $T_{\wsf_{i,p,s}}(j)$ to aggregation $\asf_{i,j}$ for each $j\in [t/2]$.

The   number of bits transmitted
through the  link from edge $\esf_{i,p}$ to aggregation $\asf_{i,j}$ is  
 \begin{align}
 & \sum_{s\in [t/2]}   |T_{\wsf_{i,p,s}}(j)|    =  \sum_{s\in [t/2]}  \frac{2|T_{\wsf_{i,p,s}}|}{t}  \nonumber\\
 & \leq   \frac{ L^{\star}(r) \Qsf \Nsf \Tsf}{\Ksf }  .\label{eq:up step 2 load}
\end{align} 

 \item We then focus on aggregation $\asf_{i,j}$ where $j\in [t/2]$. For each $p\in [t/2]$ and $s\in [t/2]$,  aggregation $\asf_{i,j}$ further divides $T_{\wsf_{i,p,s}}(j)$ into $t/2$ non-overlapping and equal-length pieces, denoted by $T_{\wsf_{i,p,s}}(j,1),\ldots,T_{\wsf_{i,p,s}}(j,t/2)$. For each $d\in [t/2]$, We  also define  $T^{\Sc}_{\wsf_{i,p,s}}(j,d)$ as the set of bits  in $T_{\wsf_{i,p,s}}(j,d)$ which are uniquely useful to servers in $\Sc$. 
The above partition  of $T_{\wsf_{i,p,s}}(j)$  is also symmetric, such that   
$$
\left|T^{\Sc}_{\wsf_{i,p,s}}(j,1)\right|=\cdots=\left|T^{\Sc}_{\wsf_{i,p,s}}(j,t/2)\right|= \frac{2|T^{\Sc}_{\wsf_{i,p,s}}(j)|}{t} ,
$$ 
for each $\Sc \subseteq [\Ksf]\setminus \{\wsf_{i,p,s}\}$.
Aggregation  $\asf_{i,j}$ sends $T_{\wsf_{i,p,s}}(j,d)$ to core $\csf_{\frac{(j-1)t}{2}+d}$ for each $d\in [t/2]$.

The   number of bits transmitted
through the  link from aggregation $\asf_{i,j}$ to core  $\csf_{\frac{(j-1)t}{2}+d}$ is  
 \begin{align}
&\sum_{p\in [t/2]} \sum_{s\in [t/2]}   |T_{\wsf_{i,p,s}}(j,d)|  =  \sum_{p\in [t/2]} \sum_{s\in [t/2]} \frac{ 2|T_{\wsf_{i,p,s}}(j)|}{t}  \nonumber\\
& =  \sum_{p\in [t/2]} \sum_{s\in [t/2]}  \frac{4|T_{\wsf_{i,p,s}}|}{t^2} \nonumber\\
& \leq  \frac{ L^{\star}(r) \Qsf \Nsf \Tsf}{\Ksf }  .\label{eq:up step 3 load}
\end{align} 
 \end{itemize}

Before introducing the downlink transmission,  
 for each pod $i\in [t]$, we define $\Nc_{i}=\{ \wsf_{i,p,s}: p\in [t/2], s\in[t/2] \},$
%\begin{align}
%\Nc_{i}=\{ k_{i,p,s}: p\in [t/2], s\in[t/2] \},
%\end{align} 
 as the set of servers connected to the edges in pod $i$.

 {\it  Downlink transmission for pod $i\in [t]$.}
 \begin{itemize}
 \item We focus on aggregation $\asf_{i,j}$ where $j\in [t/2]$. For each $d\in [t/2]$, core $\csf_{\frac{(j-1)t}{2}+d}$ sends to aggregation $\asf_{i,j}$,
 \begin{align}
 \Big\{T^{\Sc}_k(j,d): k \in [\Ksf] \setminus \Nc_i,  \Sc \subseteq [\Ksf]\setminus \{k\}, \Sc \cap \Nc_i \neq \emptyset   \Big\}.\nonumber%\label{eq:first step downlink} 
 \end{align}
Notice that  the aggregations  in pod $i$ have already received the bits in $T_k$ for $k\in \Nc_i$  from the edges in this pod, and thus  the aggregations  
need not to receive those bits from the cores.
 
The   number of bits transmitted
through the  link from core $\csf_{\frac{(j-1)t}{2}+d}$ to aggregation $\asf_{i,j}$  is  no more than
\begin{align}
%& \left|\Big\{T^{\Sc}_k(j,d):   \Sc \subseteq [\Ksf]\setminus \{k\}, \Sc \cap \Nc_i \neq \emptyset   \Big\} \right| \nonumber\\ 
& \sum_{u \in \Nc_i}  \left|\Big\{T^{\Sc}_k(j,d):  k\in [\Ksf]\setminus \{u\}, \Sc \subseteq [\Ksf]\setminus \{k\},  u \in   \Sc \Big\} \right|\nonumber\\
&  = \sum_{u \in \Nc_i}  \frac{4}{t^2} \left|\Big\{T^{\Sc}_k:   k\in [\Ksf]\setminus \{u\},  \Sc \subseteq [\Ksf]\setminus \{k\},  u \in \Sc \Big\} \right| \nonumber\\
&\leq \left(\Nsf-\frac{r\Nsf}{\Ksf } \right)  \frac{\Qsf\ssf\Tsf}{\Ksf}  .\label{eq:first step downlink load}
\end{align}
\item We then focus on edge $\esf_{i,p}$ where $p\in [t/2]$.  For each $j\in [t/2]$, the messages from the aggregation  $\asf_{i,j}$ to edge $\esf_{i,p}$ are given by
\begin{align}
 &\Big\{T^{\Sc}_k(j): k  \in [\Ksf] \setminus \{\wsf_{i,p,1},\ldots,\wsf_{i,p,t/2} \},  \Sc \subseteq [\Ksf]\setminus \{k\}, \nonumber\\& \Sc \cap \{\wsf_{i,p,1},\ldots,\wsf_{i,p,t/2} \} \neq \emptyset   \Big\}.\label{eq:second step downlink} 
\end{align}
Notice that  edge $\esf_{i,p}$ have already received   $T_k$ for $k\in \{\wsf_{i,p,1},\ldots,\wsf_{i,p,t/2} \}$  from its connected servers, and thus  edge $\esf_{i,p}$
needs not to receive those bits from the aggregations.

 The   number of bits transmitted
through the  link from aggregation  $\asf_{i,j}$    to edge $\esf_{i,p}$  is  no more than
\begin{eqnarray}
&\sum_{ u\in  \{\wsf_{i,p,1},  \ldots,\wsf_{i,p,t/2} \}}  &   \left|\Big\{T^{\Sc}_k(j): k \in [\Ksf] \setminus \{u\},  \right. \nonumber\\&& \left. \Sc \subseteq [\Ksf]\setminus \{k\}, u \in \Sc  \Big\} \right|\nonumber\\
&= \sum_{ u\in  \{\wsf_{i,p,1},  \ldots,\wsf_{i,p,t/2} \}}   &  \frac{2}{t}  \left|\Big\{T^{\Sc}_k: k \in [\Ksf] \setminus  \{u\}, \right. \nonumber\\&& \left.  \Sc \subseteq [\Ksf] \setminus  \{k\}, u \in \Sc    \Big\} \right| \nonumber\\
&\leq \left(\Nsf- \frac{ r\Nsf}{\Ksf} \right) \frac{\Qsf\ssf\Tsf}{\Ksf}   .\label{eq:second step downlink load}
\end{eqnarray}
\item Finally we focus on server $\wsf_{i,p,s}$ where  $p\in [t/2]$ and  $s\in [t/2]$. Edge $\esf_{i,p}$  sends to server $\wsf_{i,p,s}$, 
\begin{align}
U_{\wsf_{i,p,s}}  &=  \Big\{T^{\Sc}_k   : k \in [\Ksf]   \setminus  \{\wsf_{i,p,s}\}, \nonumber\\& \Sc \subseteq [\Ksf]  \setminus   \{k\}, \wsf_{i,p,s}\in \Sc    \Big\}. \nonumber%\label{eq:third step downlink} 
\end{align}
 The   number of bits transmitted
through the  link from  edge $\esf_{i,p}$   to server $\wsf_{i,p,s}$  is  no more than
\begin{align}
\left(\Nsf- \frac{  r\Nsf}{\Ksf}\right)  \frac{\Qsf\ssf\Tsf}{\Ksf}.\label{eq:third step downlink load}
\end{align}
 
 \end{itemize}

 By summing~\eqref{eq:up step 1 load} and~\eqref{eq:first step downlink load}, summing~\eqref{eq:up step 2 load} and~\eqref{eq:second step downlink load}, summing~\eqref{eq:up step 3 load} and~\eqref{eq:third step downlink load}, it can be seen that the  total number of bits transmitted through each link in the $t$-ary fat-tree is no more than 
 $$
 \frac{L^{\star}(r) \Qsf \Nsf \Tsf}{\Ksf}  +\left(\Nsf-\frac{r\Nsf}{\Ksf} \right) \frac{\Qsf\ssf\Tsf}{\Ksf}.
 $$
  Hence, we prove that the proposed scheme through 
 the $t$-ary fat-tree can achieve the optimal communication load in~\eqref{eq:CDC D star}. 
 
%{\RED Note that the links between the servers and the edges are commonly with the lowest throughput denoted by $c_{ce}$. Hence, the communication delay of the network is given by $D(r)/c_{se}$.} 

 \subsection{Discussions} 
 
%After proving that the fat-tree network can achieve the optimal max-link communication load with a low construction cost,   we will also discuss about two other practical issues. 

In this section, we will discuss two other practical issues. 

\paragraph*{Congestion at higher links}
In   hierarchical computing networks, the traffic at the top layers is always higher than the bottom layers and   the higher links always need to have higher capacities to avoid the congestion. However, in our $t$-ary fat-tree coded computing system, as we proved in Section~\ref{sub:computing with fat tree}, all links in the fat-tree topology have similar link loads (the traffic at the bottom layers is slightly higher than the top layers) and thus we can build the topology by the links with the same capacity, which also reduces the construction cost. 

\paragraph*{Fault-tolerance}
On the one hand,  any failure happening at the  switches or the links among the  switches can be tolerated, because from each server to another there are multiple paths (each edge is connected  $t/2$ aggregations and each aggregation is connected to $t/2$ cores). One the other hand, if a server node or the link connected to it  fails, we can add some redundancies in the system by using some form of the Minimum Distance Separable (MDS) code as used in \cite{d2dcaching}. However, the detailed description of this scheme is beyond the scope of this paper. 
%encode the computed intermediate value  by an Minimum Distance Separable (MDS) code, which is widely used to deal with the stragglers in the distributed computing problem. 
 
% In this paper, we considered the topological coded distributed computing problem. We first characterized the optimal max-link communication load by using the single-switch topology. To reduce the cost to build the topology, we then considered the $t$-ary fat-tree topology. In addition, we also proposed a coded distributed computing scheme through the $t$-ary fat tree, which can achieve the optimal max-link communication load. The proposed scheme can be easily extended to other computing problems with distributed servers, such as decentralized data shuffling~\cite{decentralizedDS2019wan}.

 \section{Conclusions}
\label{sec:conclusion}
 In this paper, we considered the topological coded distributed computing problem. We first characterized the optimal max-link communication load by using the single-switch topology. To reduce the cost to build the topology, we then considered the $t$-ary fat-tree topology. In addition, we also proposed a coded distributed computing scheme through the $t$-ary fat tree, which can achieve the optimal max-link communication load. The proposed scheme can avoid the congestion at higher links and tolerate the failures in the network components, such that our result has both
a significant intellectual merit and a high practical relevance.
Finally, the proposed scheme can be easily extended to other computing problems with distributed servers, such as decentralized data shuffling~\cite{decentralizedDS2019wan}.
 
\bibliographystyle{IEEEtran}
\bibliography{IEEEabrv,IEEEexample}

\end{document}